\documentclass[%
 reprint,
 aps,prl,
 twocolumn,superscriptaddress,notitlepage,nofootinbib,
 nolongbibliography
]{revtex4-2}

\usepackage{graphicx}
\usepackage{booktabs}
\usepackage{xcolor}
\usepackage{array}
\usepackage{physics}
\usepackage{dcolumn}% Align table columns on decimal point
\usepackage{bm}% bold math
\usepackage{multirow}
\usepackage{feynmp-auto}

\usepackage[colorlinks,
            linkcolor=blue,
            anchorcolor=blue,
            citecolor=blue,
            urlcolor=blue]{hyperref}

\graphicspath{{fig/}}

\def\as{\alpha_s}

\def\mbms{\overline{m}_b}
\def\mcms{\overline{m}_c}

\newcommand{\asfromBpm}{0.244}
\newcommand{\asErrfromBpm}{0.013}

\newcommand{\asfromBzero}{0.246}
\newcommand{\asErrfromBzero}{0.013}

\newcommand{\asfromB}{0.245}
\newcommand{\asErrfromB}{0.009}

\newcommand{\asValMz}{0.1266}
\newcommand{\asErrMz}{0.0023}

\newcommand{\asErrMzExp}{0.0018}

\begin{document}

\noindent
\begin{flushright}
ZU-TH 62/24 \\
\end{flushright}

\title{Determination of the strong coupling constant \texorpdfstring{$\as$}{as} from inclusive semileptonic $B$ meson decays}
% ===========================================================
\author{Yuzhi Che}
\affiliation{
  Institute of High Energy Physics, Chinese Academy of Sciences,\\
  19B Yuquan Road, Shijingshan District, Beijing 100049, China
}
\affiliation{%
 China Center of Advanced Science and Technology,
 Beijing, 100190, Beijing, China
}%

\author{Long Chen}
\affiliation{
  School of Physics, Shandong University,
  Jinan 250100, China
}

\author{Jinfei Wu}%
\affiliation{
  Institute of High Energy Physics, Chinese Academy of Sciences,\\
  19B Yuquan Road, Shijingshan District, Beijing 100049, China
}
\affiliation{%
 China Center of Advanced Science and Technology,
 Beijing, 100190, Beijing, China
}%

\author{Xinchou Lou}
\affiliation{
  Institute of High Energy Physics, Chinese Academy of Sciences,\\
  19B Yuquan Road, Shijingshan District, Beijing 100049, China
}
\affiliation{%
 University of Texas at Dallas,  Richardson, 75083, Texas, USA
}%
\affiliation{%
 Center for High Energy Physics, Henan Academy of Sciences,
 Zhengzhou, 450046, Henan, China
}

\author{Xiang Chen}
\affiliation{
  Physik-Institut, Universität Zürich,
  Winterthurerstrasse 190, CH-8057 Zürich, Switzerland
}

\author{Xin Guan}
\affiliation{SLAC National Accelerator Laboratory, Stanford University,  Stanford, CA 94039, USA}

\author{Yan-Qing Ma}
\affiliation{
  School of Physics, Peking University,   
  Beijing 100871, China
}\affiliation{
  Center for High Energy Physics, Peking University,
  Beijing 100871, China
}

\author{Manqi Ruan}
\email{ruanmq@ihep.ac.cn}
\affiliation{
  Institute of High Energy Physics, Chinese Academy of Sciences,\\
  19B Yuquan Road, Shijingshan District, Beijing 100049, China
}
\affiliation{
  University of Chinese Academy of Sciences,\\
  19A Yuquan Road, Shijingshan District, Beijing 100049, China
}

% ===========================================================
\date{\today}% It is always \today, today,

\begin{abstract}

  We demonstrate the feasibility of determining the strong coupling constant, $\as$, from the inclusive semileptonic decay width of $B$ mesons.
  We express the semileptonic $B$ decay width as a function of $\as(5\mathrm{\,GeV})$, the Cabibbo-Kobayashi-Maskawa matrix element $|V_{cb}|$, $b$- and $c$-quark masses in the $\overline{\mathrm{MS}}$ scheme.
  We fit $\as(5\mathrm{\,GeV})$ to current world averages of the $B^{\pm}$ and $B^{0}$ semileptonic decay widths.
  This yields $\as(5\mathrm{\,GeV}) = \asfromB \pm \asErrfromB$, corresponding to a 5-flavor extrapolation of $\as(m_{Z}) = \asValMz \pm \asErrMz$.
  The primary uncertainty contributions arise from the uncertainty on the perturbative expansion and the value of $|V_{cb}|$.
  Future advancements including higher-order perturbative calculations,
  and precise measurements of $|V_{cb}|$ and $B$ decay widths from upcoming $B$ and $Z$ factories,
  could enable this method to determine $\as(m_{Z})$ with a competitive precision of $\Delta\as(m_{Z}) \sim \asErrMzExp$.
  This precision is comparable to the current accuracy of $\as(m_{Z})$ measurements from $\tau$-lepton decays, which is regarded as the most precise experimental approach.

\end{abstract}

\maketitle
% ===========================================================

\section{Introduction}

The strong interaction, one of the fundamental interactions in nature,
is described by quantum chromodynamics (QCD). The strong coupling
constant, $\as(\mu)$, characterizes the strength of this interaction
and exhibits a decreasing trend with increasing energy scale $\mu$.
This running behavior is described by the renormalization group
equation (RGE)~\cite{prosperi2007running}, reflecting essential
properties of the strong interaction, such as quark confinement at long
distances and asymptotic freedom at short distances. Consequently,
precise knowledge of $\as(\mu)$ across the entire range of energy scale
is crucial for a comprehensive understanding and testing of QCD. $\as$
at low energy scale has been studied through various methodologies,
including hadron production in electron-positron
annihilation~\cite{boito2018Stronga}, semileptonic charmed
meson~\cite{Wu:2024jyf} and $\tau$ decays~\cite{Braaten:1991qm,
  cleocollaboration1995Measurement, theopalcollaboration1999Measurement,
  davierUpdateALEPHNonstrange2014, boito2018Stronga, Boito:2020xli,
  Baikov:2008jh}, and inclusive hadronic decay of heavy
quarkonia~\cite{shen2023Novel, Narison:2018xbj}. However, there are
relatively few measurements of $\as$ in the energy scale range around
$5\mathrm{\,GeV}$.

\begin{figure}[h]
  \centering
  \includegraphics[width=0.45\linewidth]{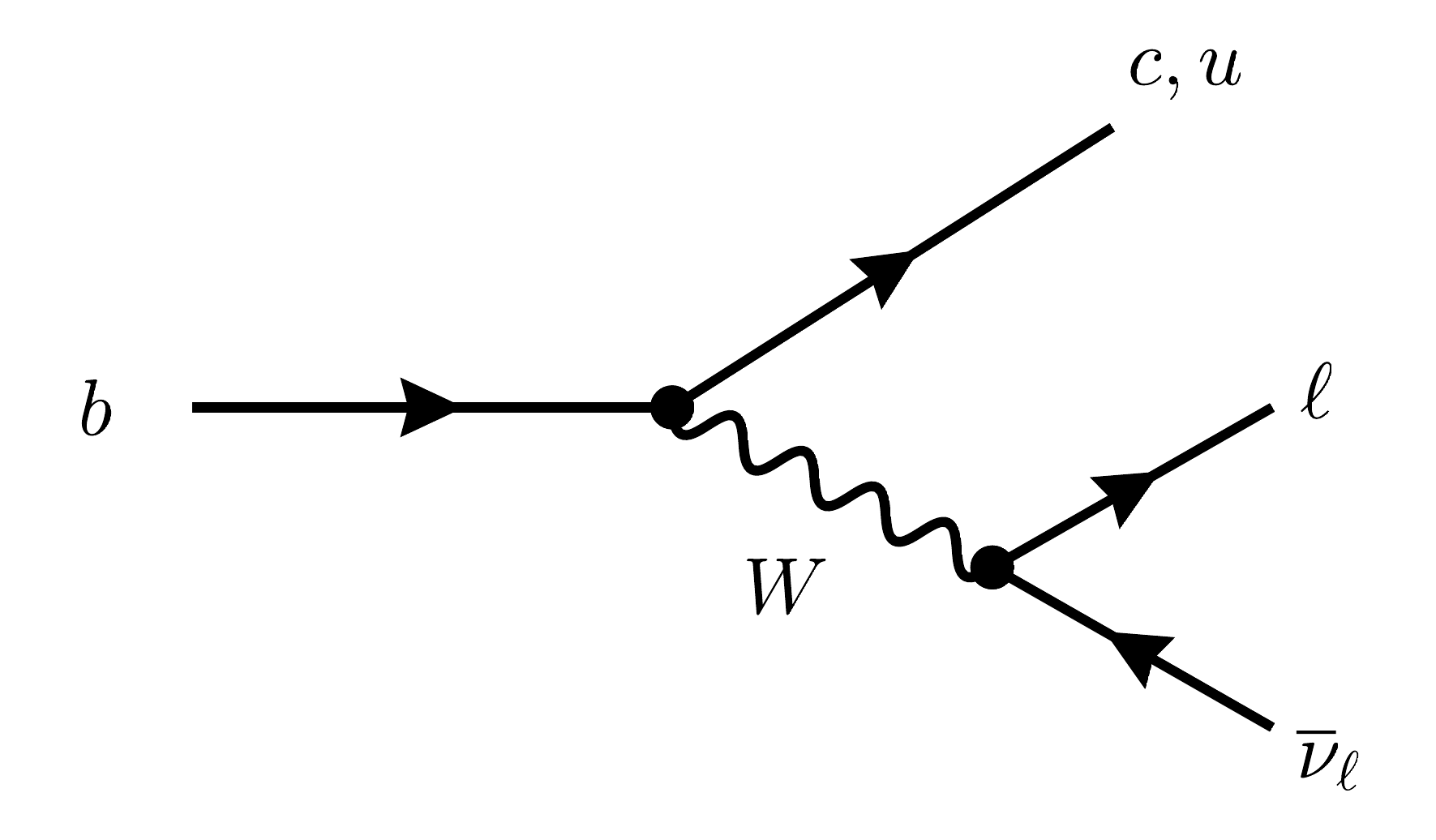}
  \caption{\label{Feynman} The parton level Feynman diagram of semileptonic $B$ decay.}
\end{figure}

We consider measuring $\as$ from the inclusive semileptonic $B$ decay
which corresponds to the energy scale of $B$ meson masses.
Figure~\ref{Feynman} shows the Feynman diagram for the inclusive
semileptonic $B$ decay ($B \to X \ell\nu$) at the tree level in the
parton model. This process consists of two components: $B\to X_{c}
  \ell\nu$ and $B\to X_{u} \ell\nu$, where $X_{c}$ represents the charmed
system and $X_{u}$ the light hadron system. The ratio of $B\to X_{u}
  \ell\nu$ is approximately 65 times less than the former due to Cabibbo
suppression. Using the Heavy Quark Expansion (HQE) method, the
branching ratio and the spectral moments of kinematic observables have
been parameterized as functions of the Cabibbo-Kobayashi-Maskawa (CKM)
matrix elements, $|V_{cb}|$ and $|V_{ub}|$, the strong coupling
constant ($\as$), $b$-quark mass ($m_{b}$), $c$-quark mass ($m_{c}$)
and non-perturbative HQE parameters~\cite{Gambino:2004qm,
  benson2003Imprecated,
  dowling2008Semileptonic,Pak:2008qt,Pak:2008cp,Melnikov:2008qs,Gambino:2011cq,
  fael2019Vcb, PhysRevD.104.016003,fael2024NNLO}. On the experimental
side, these observables have been measured by the BaBar, Belle, and
Belle II collaborations over the past two
decades~\cite{brandt2002Semileptonic, aubert2010Measurementa,
  Belle:2006kgy, collaboration2021Measurement}.

The $B\to X_c \ell\nu$ process was used to determine $|V_{cb}|$,
$m_{b}$ and $m_{c}$, with $\as$ fixed at the value extrapolated from
the world average of $\as(m_{Z})$~\cite{Gambino:2013rza, fael2019Vcb,
  gambino2015Inclusive, bordone2021Threea}. Nowadays, more precise
determinations of the $|V_{cb}|$, $m_{b}$ and $m_{c}$ are available,
for example, $|V_{cb}|$ from exclusive $B$
decays~\cite{Grinstein:2017nlq, FermilabLattice:2021cdg, Gao:2021sav,
  Cui:2023jiw, Belle:2023xgj} or $W$ decays~\cite{liang2024Measurement,
  aidanrichardwiederhold2024Flavour}; heavy quark masses from lattice
QCD~\cite{hatton2021Determination, bazavov2018strange,
  alexandrou2021Quark, hatton2020Charmonium, heitger2021Determination},
$b$- and $c$-meson masses~\cite{Kiyo:2015ufa, Ayala:2016sdn,
  narison2020overline, peset2018charm}, or $e^+e^- \to \text{hadrons}$
cross-section~\cite{Hoang:1999ye, Melnikov:1998ug, Beneke:1999fe,
  Dehnadi:2011gc, Chetyrkin:2017lif}, etc. Therefore, by fixing the
values of $|V_{cb}|$, $m_{b}$ and $m_{c}$ according to those
progresses, we could extract $\as$ using the semileptonic $B$ decay
width at the scale around the $B$ meson masses.

\section{Theoretical model\label{sec:theory}}

\begin{table*}
  \caption{The parameters used during the construction of the theoretical model.}
  \label{tab:param}
    \begin{tabular}{@{} lllc}
      \toprule
      Parameter                                & Notation             & Value \& error                                 & Note                     \\
      \midrule
      Fermi coupling constant                  & $G_{F}$              & $1.16637886 \times 10^{-5} \mathrm{~GeV^{-2}}$ & \cite{pdg}               \\
      Electroweak correction factor            & $A_{\mathrm{ew}}$    & $1.014$                                        & \cite{Sirlin:1981ie}     \\
      CKM matrix element                       & $|V_{cb}|$           & $0.0398 \pm 0.0006$                            & \cite{pdg}               \\
      $b$-quark mass in $\overline{\text{MS}}$ & $\mbms(\mbms)$       & $4.18 ^{ +0.03 }_{- 0.02 } \mathrm{\, GeV}$    & \cite{pdg}               \\
      $c$-quark mass in $\overline{\text{MS}}$ & $\mcms(\mcms)$       & $1.27 \pm 0.02 \mathrm{\, GeV}$                & \cite{pdg}               \\
      \multirow{4}{*}{HQE parameters}          & $\mu_{\pi}^2$        & $0.477 \pm 0.056 \mathrm{\,GeV^2}$             & \cite{bordone2021Threea} \\
                                               & $\mu_{G}^2$          & $0.306 \pm 0.050 \mathrm{\,GeV^2}$             & \cite{bordone2021Threea} \\
                                               & $\rho_{D}^3$         & $0.185 \pm 0.031 \mathrm{\,GeV^2}$             & \cite{bordone2021Threea} \\
                                               & $\rho_{LS}^3$        & $-0.130 \pm 0.092 \mathrm{\,GeV^2}$            & \cite{bordone2021Threea} \\
      $b$-quark mass in kinetic scheme         & $m_{b}^{\text{kin}}$ & $4.573 \pm 0.012 \mathrm{\,GeV}$               & \cite{bordone2021Threea} \\
      \bottomrule
    \end{tabular}
\end{table*}

Within the Heavy Quark Expansion (HQE) framework, the inclusive
semileptonic $B$ decay width takes the form shown in~\eqref{eq:gammaB}.
\begin{equation}
  \begin{aligned}
  \Gamma\left(B \rightarrow X_c \ell \bar{\nu}_{\ell}\right)=\Gamma_0[ & C_0-C_{\mu_\pi} \frac{\mu_\pi^2}{2 m_b^2}+C_{\mu_G} \frac{\mu_G^2}{2 m_b^2}\\
  & -C_{\rho_D} \frac{\rho_D^3}{2 m_b^3}-C_{\rho_{L S}} \frac{\rho_{L S}^3}{2 m_b^3}+\cdots],
  \end{aligned}
  \label{eq:gammaB}
\end{equation}

where $\Gamma_0 \equiv \frac{G^2_F |V_{cb}|^2 m_b^5 A_{\mathrm{ew}}}{192 \pi^3}$,
with $G_F = 1.16637886 \times 10^{-5} \mathrm{~GeV^{-2}}$ is the Fermi coupling constant, and $A_{\mathrm{ew}} = 1.014$ is the electroweak correction factor~\cite{Sirlin:1981ie}.
The coefficients $C_{i}$ ($i=0, \mu_\pi, \mu_G$) depend on the ratio of the squared $c$- and $b$-quark masses, $\rho = \frac{m_c^2}{m_b^2}$, and have perturbative expansions in $\as$.
The $\mu_{\pi}^2$, $\mu_{G}^2$, $\rho_D^3$ and $\rho_{L S}^3$ are the HQE parameters corresponding to the kinetic, chromomagnetic, Darwin and spin-orbital terms, respectively.
For the purpose of determining $\as$, a numerical mapping between $ \Gamma\left(B \rightarrow X_c \ell \bar{\nu}_{\ell}\right)$ and $\as$ has been established,
considering the leading order power correction up to the fourth order, as well as the leading order contributions from the four terms of high order power corrections.
The precision of the HQE parameters, the next-to-leading order contributions of the high order power corrections, and the contributions of order $\mathcal{O}(1/m_b^{4,5})$ have been included in the error.

The perturbation expansion of $C_0$ can be expressed as: $C_0 =
  \mathbf{c}_0 + \mathbf{c}_1 \frac{\as}{\pi} + \mathbf{c}_2
  \Big(\frac{\as}{\pi}\Big)^2 + \mathbf{c}_3 \Big(\frac{\as}{\pi}\Big)^3
  + \mathcal{O}(\as^4)$, where the leading term $\mathbf{c}_0 = 1 - 8\rho
  + 8\rho^3 - \rho^4 - 12\rho^2\operatorname{ln} \rho$ is the tree-level
phase space factor~\cite{benson2003Imprecated}. The results for the
second-order~\cite{Pak:2008qt,Pak:2008cp,Melnikov:2008qs} and
third-order~\cite{bordone2021Threea,PhysRevD.104.016003} perturbative
corrections have been provided in the on-shell scheme. For better
perturbative convergence, these perturbative QCD results are
reformulated in the $\overline{\text{MS}}$ scheme, at the
renormalization scale $\mu = 5~\mathrm{GeV}$ the mass scale of the
decaying $B$ meson. For the purpose of extracting
$\as(5\mathrm{\,GeV})$ in 5-flavor scheme, the perturbative correction
$C_0$ is reformulated consistently in terms of the
$\overline{\text{MS}}$-renormalized quark masses $\mbms(\mu)$,
$\mcms(\mu)$ and $\as(\mu)$. The values for the arguments
$\mbms(5\mathrm{\,GeV})$ and $\mcms(5\mathrm{\,GeV})$ (in 5-flavor
scheme) are derived by solving the RGE system with the boundary
conditions for $\mbms(\mu), \mcms(\mu)$ and $\as(\mu)$ set as
following: the Particle Data Group (PDG) average values of
$\mcms(\mcms) = 1.27 \pm 0.02 \mathrm{\,GeV}$ and $\mbms(\mbms) =
  4.18^{+0.03}_{-0.02} \mathrm{\,GeV}$~\cite{pdg}, and the sampled values
of $\as(5\mathrm{\,GeV})$ in the fit. In this way the perturbative
correction to $\Gamma\left(B \rightarrow X_c \ell
  \bar{\nu}_{\ell}\right)$ up to $\mathcal{O}(\alpha_s^3)$ in the
leading-power correction is eventually expressed as a numerical
function of $\as(5\mathrm{\,GeV})$.

\begin{figure}[h]
  \centering
  \includegraphics[width=0.4\textwidth]{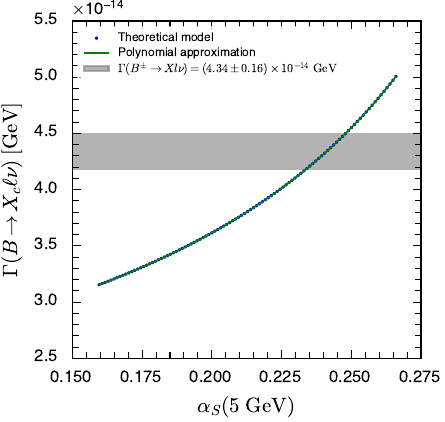}
  \caption{The numerical function of $\Gamma(B\to X_{c}\ell \nu)$
    versus $\as(5\mathrm{\,GeV})$, compared with the $\Gamma(B^{\pm}\to X_{c} \ell\nu) = (4.34\pm 0.16)\times 10^{-14} \mathrm{\,GeV}$
    derived from Eq.~\eqref{eq:decay_width}. The numerical function is parameterized by a polynomial function in the $\as(5\mathrm{\,GeV})$ range from $0.16$ to $0.26$.}
  \label{fig:Gamma_as}
\end{figure}

The numerical calculations for the coefficients $C_{\mu_{\pi}}$ and
$C_{\mu_{G}}$ were detailed in~\cite{Alberti:2013kxa}. For the
$1/m_b^3$ power corrections, the coefficients related to the Darwin
term and spin-orbital were discussed in~\cite{Gremm:1996df,
  Mannel:2019qel}. The HQE parameters $\mu_{\pi}^2$, $\mu_{G}^2$,
$\rho_{D}^3$, and $\rho_{LS}^3$ have been measured through a
simultaneous fit using the spectral moments of semileptonic $B$
decays~\cite{bordone2021Threea}. In addition, the $\mu_{\pi}^2$ and
$\mu_{G}^2$ can also be derived from the mass splitting of $B$
mesons~\cite{Gambino:2017dfa}. The determination of $\rho_{D}^3$ is
expected to be achieved with high precision using $D$ decay data from
BESIII~\cite{Bernlochner:2024vhg}. Using leading-order approximations of $1/m_b^{2,3}$ power
corrections (Eq.~(4.1) in~\cite{Alberti:2013kxa} and Eq.~(23)
in~\cite{Gremm:1996df}) and parameters listed in Table~\ref{tab:param},
we estimate the high order power corrections to decrease the decay
width by $\sim 7\%$, relative to $\Gamma_0\mathbf{c}_0$.~\footnote{It
  is consistent with the numerical values in Eq.(4.1)
  of~\cite{Mannel:2010wj}.}

Overall, the numerical mapping from the assumed values of
$\as(5\mathrm{\,GeV})$ to the decay width $\Gamma(B\to X_c \ell\nu)$ is
shown in Fig.~\ref{fig:Gamma_as}. The involved parameters are
summarized in Table~\ref{tab:param}. Within the range of
$\as(5\mathrm{\,GeV}) \in [0.16,\, 0.26]$, the decay width dependence
is parameterized by a fifth-order polynomial: $\Gamma\left(B
  \rightarrow X_c \ell \bar{\nu}_{\ell}\right) = (1.01\times
  10^{-14}\mathrm{\,GeV})\, \big(-1 + 88.9\, \alpha_s^{\mathrm{fit}} -
  904.8\, (\alpha_s^{\mathrm{fit}})^2 + 4946.4\,
  (\alpha_s^{\mathrm{fit}})^3 - 13467.1\, (\alpha_s^{\mathrm{fit}})^4 +
  15502.7\, (\alpha_s^{\mathrm{fit}})^5 \big)$ where
$\alpha_s^{\mathrm{fit}} \equiv \as(5\mathrm{\,GeV})$.

\begin{table*}
  \centering
  \caption{The relative uncertainty contributions to the theoretical prediction of $\Gamma_{sl}$ and the $\as(5\mathrm{\,GeV})$ fitting result using $\Gamma(B^{\pm} \to X_{c} \ell\nu)$.
  Values in the parenthesis are the perspective values considering future improvements.
  The $V_{cb}$ will be measured from $W$ decays~\cite{liang2024Measurement}.
  The uncertainty of the branching ratio will be reduced by a factor of about 0.05 considering the $50\mathrm{\,ab^{-1}}$ data to be collected by Belle II~\cite{Belle-II:2018jsg}.
  The accuracy of $b$- and $c$-quark masses has been achieved by lattice QCD calculation~\cite{pdg}.
  The R-scale $\mu$ uncertainty will be further controlled by higher-order perturbative calculations.
  See body text for more details.}
  \label{tab:err_divid}
    \begin{tabular}{@{} lccc}
      \toprule
                                                                                       & $\Gamma_{sl}$ prediction $[\%]$ & $\as(5\mathrm{\,GeV})$ $[\%]$ \\
      \midrule
      $|V_{cb}| = 0.0398 \pm 0.0006$                                                   & 3.0~(1.4)                       & 2.1~(1.0)                     \\
      $\overline{m}_{b}(\overline{m}_{b}) = 4.18 ^{ +0.03 }_{- 0.02 } \mathrm{\, GeV}$ & 3.0~(1.1)                       & 2.1~(0.8)                     \\
      $\overline{m}_{c}(\overline{m}_{c}) = 1.27 \pm 0.02 \mathrm{\, GeV}$             & 2.1~(1.4)                       & 1.4~(1.0)                     \\
      R-scale $\mu = 5^{+5}_{-2.5} \mathrm{\,GeV}$                                     & 4.4~(2.2)                       & 3.1~(1.6)                     \\
      High-order power corrections                                                     & 2.3~(2.3)                       & 1.6~(1.6)                     \\
      $\tau_{B^{\pm}} = 1.638 \pm 0.004 \mathrm{\,ps}$                                 & -                               & 0.2                           \\
      $\mathcal{B}(B^{\pm} \to X_{c} \ell\nu) = 10.8 \pm 0.4 ~\%$                      & -                               & 3.0~(2.2)                     \\
      \midrule
      Sum                                                                              & 6.9(3.2)                        & 5.7~(3.5)                     \\
      \bottomrule
    \end{tabular}
\end{table*}

\section{Result and discussion\label{sec:result}}
We fit the value of $\as(5\mathrm{\,GeV})$ to the inclusive semileptonic decay widths of the $B^{\pm}$ and $B^{0}$ mesons.
The experimental decay width can be obtained from measured values for the lifetime and semileptonic decay branching ratio:
\begin{equation}
  \Gamma_{sl} = \frac{\hbar}{\tau} \mathcal{B}_{sl} \,.
  \label{eq:decay_width}
\end{equation}
We quote the world averages for the lifetime~\cite{pdg} and the partial branching ratios (with a cut on the lepton energy, $E_{l} > 0.4\mathrm{\,GeV}$) measured by the Belle experiment~\cite{Belle:2006kgy},
$$
  \tau_{B^{\pm}} = 1.638 \pm 0.004 \mathrm{\,ps}, \quad
  \mathcal{B}(B^{\pm} \to X_{c} \ell\nu) = 10.8 \pm 0.4 \%,
$$
$$
  \tau_{B^{0}} = 1.517 \pm 0.004 \mathrm{\,ps}, \quad
  \mathcal{B}(B^{0} \to X_{c} \ell\nu) = 10.1 \pm 0.4 \%.
$$
The total branching ratios are obtained by scaling the particle branching ratios by a factor of $1.015$, according to~\cite{Gambino:2013rza}.
As a result, we obtain $\Gamma(B^{\pm}) = (4.40 \pm 0.16) \times 10^{-14} \mathrm{\,GeV}$ and $\Gamma(B^{0}) = (4.44 \pm 0.17) \times 10^{-14} \mathrm{\,GeV}$.
For the related parameters, $|V_{cb}|$ is fixed at the PDG world average value~\cite{pdg}, which is extracted from the $\overline{B} \rightarrow D^* \ell \bar{\nu}_{\ell}$ decays (with $\ell = e, \mu$) along with the lattice QCD calculation of the form factors, independent of the perturbative $\as$.
The minimum-$\chi^2$ fit incorporates the experimental errors in the decay widths, as well as the theoretical uncertainties introduced by parameters and the numerical expression of HQE.
These uncertainties are considered to be independent of each other.

\begin{figure}[h]
  \centering
  \includegraphics[height=0.3\textwidth]{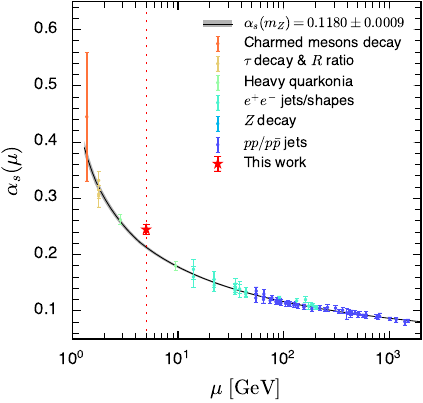}
  \includegraphics[height=0.3\textwidth]{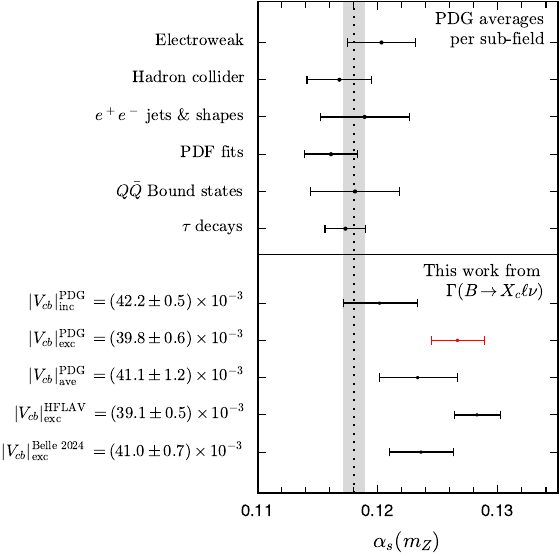}
  \caption{(Top) The combined $\as(5\mathrm{\, GeV})$ result (Eq.~\eqref{eq:res}) compared with the $\as$ measurements at other energy scales~\cite{Wu:2024jyf, Boito:2020xli, Baikov:2008jh, boito2018Stronga, Narison:2018xbj, CMS:2016lna, D0:2009wsr, D0:2012xif, ATLAS:2017qir, CMS:2014mna, Schieck:2012mp, Dissertori:2009ik, Bethke:2009ehn}.
    (Bottom) The comparison of the $\as(m_{Z})$ pre-averages from six experimental sub-fields in PDG~\cite{pdg} and the extrapolated values from this work.
    Additionally, the $\as(m_{Z})$ derived using different values of $|V_{cb}|$ are also compared for reference, including the PDG average~\cite{pdg} for inclusive determination ($|V_{cb}|^{\text{PDG}}_{\text{inc}}$), 
    exclusive determination ($|V_{cb}|^{\text{PDG}}_{\text{exc}}$), their average ($|V_{cb}|^{\text{PDG}}_{\text{ave}}$), exclusive $|V_{cb}|$ from HFLAV group ($|V_{cb}|^{\text{HFLAV}}_{\text{exc}}$)~\cite{Banerjee:2024znd},
    and recent exclusive determination from Belle ($|V_{cb}|^{\text{Belle 2024}}_{\text{exc}}$)~\cite{Belle:2023xgj}.
    The extrapolation of $\as$ along the energy scale is conducted using the \textsc{RunDec} package~\cite{chetyrkin2000RunDec}.
  }
  \label{fig:alpha_S}
\end{figure}

The fit yields $\as(5\mathrm{\,GeV}) = \asfromBpm \pm \asErrfromBpm$
from $\Gamma(B^{\pm}\to X_{c} \ell\nu)$, and $\as(5\mathrm{\,GeV}) =
  \asfromBzero \pm \asErrfromBzero$ from $\Gamma(B^{0}\to X_{c}
  \ell\nu)$. Combining the two fits gives:
\begin{equation}
  \as(5\mathrm{\,GeV})= \asfromB \pm \asErrfromB.
  \label{eq:res}
\end{equation}
The combined results are shown and compared with other determinations in Fig.~\ref{fig:alpha_S}.

The uncertainty on $\as$ arises from three sources: the experimental
measurements of the branching ratio and lifetime, the theoretical
uncertainty induced by uncertainties on the used parameters and
perturbation expansion for the coefficients of HQE. These terms of the
$\as$ uncertainty are listed in Table~\ref{tab:err_divid}, taking the
fit with $\Gamma(B^{\pm} \to X_{c} \ell\nu)$ as an example. The
uncertainty in $|V_{cb}|$ propagates to $\Gamma_{sl}$ via $\Gamma_{0}$
as follows: $$ \left.\frac{\sigma (\Gamma)}{\Gamma}\right|_{|V_{cb}|} =
  2 \frac{\sigma(|V_{cb}|)}{|V_{cb}|} $$ contributing a relative
uncertainty of $3.0\%$ on $\Gamma_{sl}$. The errors induced by the
uncertainties on the input values of
$\overline{m}_{c}(\overline{m}_{c})$ and
$\overline{m}_{b}(\overline{m}_{b})$ are estimated by taking the
largest deviations from varying their values within the respective
error bounds. The results show that uncertainties in
$\overline{m}_{c}(\overline{m}_{c})$ and
$\overline{m}_{b}(\overline{m}_{b})$ contribute about $2\%$ and $3\%$
relative uncertainty to the $\Gamma_{sl}$ prediction, respectively. The
uncertainty due to the remnant renormalization scale dependence
(R-scale uncertainty) of the leading-order power correction is
estimated by varying $\mu$ from $2.5\mathrm{\,GeV}$ to
$10\mathrm{\,GeV}$. It leads to about $-2\%$ to $-4.4\%$ variations in
the perturbative corrections relative to the result at $\mu =
  5\mathrm{\,GeV}$. The larger variation is taken as the estimation of
the uncertainty. The uncertainty of the high order power corrections
are estimated by summing up two contributions quadratically. First, the
errors on the $\mu_{\pi}^2$, $\mu_{G}^2$, $\rho_{D}^3$, $\rho_{LS}^3$,
and $m_{b}^{\text{kin}}$ are considered, leading an uncertainty of
about $0.9\%$ on the theoretical prediction for $\Gamma\left(B
  \rightarrow X_c \ell \bar{\nu}_{\ell}\right)$. The second component is
the truncation error. The next-to-leading order contributions of
$C_{\mu_\pi} \frac{\mu_\pi^2}{2 m_b^2}$, $C_{\mu_G} \frac{\mu_G^2}{2
    m_b^2}$, $C_{\rho_D} \frac{\rho_D^3}{2 m_b^3}$ are estimated according
to~\cite{Alberti:2013kxa} and~\cite{Mannel:2010wj}. By summing the
absolute values of these next-leading-order corrections, the truncation
error for $1/m_b^{2,3}$ power corrections are estimated conservatively.
Meanwhile, the $\mathrm{O}(1/m_b^{4,5})$ order power corrections are
evaluated to cause a reverse influence by a factor of $1.3\%$
in~\cite{Mannel:2010wj}, which is also added into the total truncation
error. As a result, we assign a truncation error of $2.3\%$ on the high
order power corrections. \footnote{During the estimation of each order
  power correction, results from different schemes are directly used
  without converting them to a consistent scheme. The effect of this
  approach is on the next-to-leading order, and included in the
  uncertainty estimation.} The experimental uncertainty of the branching
ratio and life-time also contributes to $\as(5\mathrm{\, GeV})$
uncertainty.

The result of $\as$ fit in~\eqref{eq:res} is extrapolated to the scale
of $m_Z$, $\as(m_{Z}) = \asValMz \pm \asErrMz$. As shown in the second
plot of Fig.~\ref{fig:alpha_S}, the equivalent $\as(m_{Z})$ from this
study exhibits accuracy comparable to the PDG pre-averages from other
$\as$ determination fields. The primary source of the uncertainty
arises from the RG-scale uncertainty, which will be refined by future
perturbative calculations. Based on the observed reduction in the
conventional perturbative QCD scale uncertainty from
$\mathcal{O}(\alpha_s^2)$ to $\mathcal{O}(\alpha_s^3)$ for $b
  \rightarrow c \ell\nu$ and $b \rightarrow u \ell\nu$, it is plausible
to anticipate that the knowledge of the next order result may halve
this perturbative uncertainty. The recent lattice QCD results have
determined the quark mass with uncertainties around the
$10\mathrm{\,MeV}$ level and further improvements are anticipated. The
$|V_{cb}|$ measurements from $W$ boson decays are expected to achieve
the accuracy of $\sim 0.7\%$ on the future electron-positron
collider~\cite{liang2024Measurement,
  aidanrichardwiederhold2024Flavour}. The current measurements of the
semileptonic $B$ decay branching ratios are derived from the
$140\mathrm{\,fb^{-1}}$ of data collected by
Belle~\cite{Belle:2006kgy}, with statistical and systematic
uncertainties being comparable. Among these, the statistical term will
decrease by a factor of approximately 20 when the data set increases to
$50\mathrm{\,ab^{-1}}$ on the Belle II~\cite{Belle-II:2018jsg}.
Assuming the systematic uncertainty remains at the same level, the
experimental uncertainty of $B$ decay width will be $\sim 0.3\%$. All
these improvements are scaled to the perspective uncertainties on the
$\Gamma_{sl}$ and $\as(5\mathrm{\,GeV})$, and marked in parenthesis in
Table~\ref{tab:err_divid}. Taking into account those advancements, the
$\as(m_{Z})$ determination could eventually reach $\pm \asErrMzExp$,
halving the precision conducted by this research. This precision is
comparable to the current precision of $\as$ measurement from $\tau$
decays, which is considered one of the most precise approaches.

In addition, this fit uses the exclusive determination of $|V_{cb}|$ as an external input.
However, the world average value of the exclusive $|V_{cb}|$ shows $3\sigma$ tension with the inclusive one~\cite{Banerjee:2024znd}. 
Consequently, the $\as$ determination is entangled with this $|V_{cb}|$ puzzle.
Figure~\ref{fig:alpha_S} illustrates the impact of the $|V_{cb}|$ value on
the resulting $\as(5\mathrm{\,GeV})$. 
More recently, the exclusive $|V_{cb}|$ has been determined to be $|V_{cb}| = (41.0\pm 0.7)\times 10^{-3}$ using the Belle data~\cite{Belle:2023xgj},
which is also compared in Fig.~\ref{fig:alpha_S}.

\section{Summary\label{sec:conclusion}}
This manuscript discusses the feasibility to determine $\as(5\mathrm{\,GeV})$ from the inclusive semileptonic $B$ decay width.
The theory model is based on the framework of HQE and includes the $|V_{cb}|$, $\mbms(\mbms)$, $\mcms(\mcms)$, and four HQE parameters.
By constrain the other parameters at external determinations listed in Table~\ref{tab:param},
it is possible to achieve an $\as(5\mathrm{\,GeV})$ determination of $\as = \asfromB \pm \asErrfromB$, corresponding to $\as(m_Z) = \asValMz \pm \asErrMz$.

The uncertainty is estimated to be comparable to the averages of
$\as(m_{Z})$ from other experimental methods. The main sources of the
uncertainty are estimated in Table~\ref{tab:err_divid}. With further
improvements in perturbation calculations and the measurement accuracy
of related parameters, the uncertainty of this method could be halved.

It should be noted that as a fundamental parameter, $\as$ influences
QCD predictions through multiple parameters and calculations. A
challenge in $\as$ determination is that theoretical models correlate
with prior $\as$ assumptions. To address this challenge, we use the
measurement of $|V_{cb}|$ from exclusive $B$ decays, which is, in
principle, independent of the perturbative $\as$. We also include the
$\as$ dependency of the scheme transformation and the scale evolution
of the quark masses in the $\as$ fitting. However, the values of
$\mbms(\mbms)$, $\mcms(\mcms)$ used in our present analysis are taken
from the PDG averages and the HQE parameters are determined from
simultaneous fit using spectral moments of inclusive $B\to
  X_{c}\ell\nu$ decays. These parameters depend on the assumptions of the
perturbation $\as$. To further mitigate this implicit correlation, one
potential future avenue is to reformulate the perturbative corrections
using the renormalized quark masses defined in the
regularization-independent momentum-subtraction
schemes~\cite{Martinelli:1994ty,Sturm:2009kb,Boyle:2016wis,Lytle:2018evc,DelDebbio:2024hca},
which can be directly extracted from lattice calculations independently
of the perturbative $\as$. Alternatively, it is also worth considering
a future global fitting of $\as$, $m_{b}$, and $m_{c}$ using a broader
range of observables, such as spectral moments of semileptonic $B$ and
$D$ decays, and the masses of $B$ mesons.
% ===========================================================
\begin{acknowledgments}

\emph{\textbf{Acknowledgments}---}  The authors would like to thank Yuming Wang and Lu Cao for the fruitful discussion on the topic of HQE.
This study was supported by the National Key R\&D Program of China (Grant NO.: 2022YFE0116900),
the National Natural Science Foundation of China (Grants No.~12342502, No.~12325503, No.~12321005, No.~12235008, No.~12205171, No.~12042507),
Natural Science Foundation of Shandong province under contract 2024HWYQ-005, tsqn202312052.
the United States Department of Energy, Contract DE-AC02-76SF00515,
and the Swiss National Science Foundation (SNF) under contract 200020\_219367.

\end{acknowledgments}
% ===========================================================

% ===========================================================
\bibliographystyle{apsrev4-1} 
\bibliography{biblio}

\end{document}